\documentstyle[aps,preprint,epsf]{revtex}

\begin{document}

\draft
\title{Theory of $\omega^{-4/3}$ 
law of the power spectrum in dissipative flows}
\author{Hisao Hayakawa}
%\email{hisao@yuragi.jinkan.kyoto-u.ac.jp}
%\affiliation
\address{Department of Physics, Yoshida-South Campus, Kyoto University,
Kyoto 606-8501, Japan}

\maketitle
\begin{abstract}
 It is demonstrated that $\omega^{-4/3}$ law of the power spectrum with
 the angular frequency $\omega$
 in dissipative flows is produced by the emission of dispersive
 waves from the antikink of an  congested domain. The analytic theory
 predicts the spectrum is proportional to $\omega^{-2}$ for relatively
 low frequency  and $\omega^{-4/3}$ for high frequency.
 \end{abstract}
%\pacs{81.05.Rm,47.20.-k 05.20.Dd}
Recently, much attention has been attracted to collective dynamics of
dissipative particles\cite{jaeger,goldhirsch}. 
In particular, physics of 
granular flows\cite{nishi,pouliquen,namiko02,namiko04} 
and traffic flows\cite{helbing,tgf01} are developing subjects.
In such dissipative flows, we often observe the coexistence of congested
regions and dilute regions. It is important to know the mechanism of the
emergence of congestion of traffic and granular flows.  Although
we have some exact results on the formation of congested domains in
one-dimensional traffic flow\cite{igarashi,igarashi2}, we still do not understand the details of
the fluctuation of dissipative flows. 

In experiments of dissipative flows, we usually measure the power
spectrum which is the Fourier transform of the auto-correlation
function. It is known that traffic flows  and granular flows in a pipe
have the power spectra obeying $\omega^{-\beta}$ law with the angular
frequency $\omega$\cite{horikawa,musha}.
%In particular, to know the mechanism of $1/\omega$-spectrum is still
%challenging. 
%On the other hand, we have understood the mechanism  of $f^{-3/2}$ law
%as a result of diffusive relaxation of structural materials.
Several years ago, Moriyama {\it et al}.\cite{moriyama} have confirmed that granular
flow in a pipe should have the spectrum with $\beta=4/3$. We also expect that the power
spectrum obeying 
 $\omega^{-4/3}$ law is universal for dissipative flows 
in the coexistence of congested-flow 
and sparse-flow.\cite{nishi,moriyama,peng,yamazaki,ptp150}
 This law is robust in the experiments
of granular flows, which can be observed without tuning of a  suitable set of
parameters.\cite{moriyama,ptp150}

Although the previous papers\cite{nishi,moriyama} 
proposed the mechanism of $\omega^{-4/3}$ law,
their derivation might be incomplete. We can list
several defects in their derivation: (i)
They assumed that the system is in a weakly stable region 
of homogeneous state. However,  $\omega^{-4/3}$ can be commonly 
observed  in the case of the coexistence
between congestion and the sparse-flow. The  power exponent
$\beta$ is drastically small when there are no definite domains in
systems.\cite{yamazaki,ptp150}  
(ii) The experiments\cite{moriyama,ptp150} 
suggest that $\omega^{-4/3}$ law is robust without
 fine-tuning when phase separations take place, but the theory assumes
 that the system is in the vicinity of the neutral curve of the linear
 stability analysis.
(iii) The theoretical spectrum depends on the wave number
but there is no wave-number dependence in the actual observation in
 experiments.\cite{nishi,moriyama} 
(iv) Although the theory assumes that 
the relaxation process of internal structures, it is not clear what
 the relevant relaxation process is.
Therefore, one is skeptical of the validity of the previous theory
  to explain $\omega^{-4/3}$ law.

Recently, Takesue {\it et al.}\cite{takesue} have solved a kink-diffusion problem in the totally asymmetric 
simple exclusion process (TASEP)\cite{asep} and  derived $\omega^{-3/2}$
law of the power spectrum.
Although TASEP contains only a kink  which connects one congested domain
with a dilute region,  
their analysis is suggestive to 
understand more realistic situations in  traffic and granular flows.
In this Communication, we thus try to re-derive $\omega^{-4/3}$ law
in the case of coexistence between congestion
and sparse-flow.

In order to proceed the analysis we should recall that
all of one-dimensional models for traffic and granular flows in weakly
unstable regions can
be described by trains of 
quasi-solitons stabilized by small dissipations.\cite{nishi,komatsu,hn98,wada} 
In general, 
a dilute region is connected with a congested region by asymmetric
interfaces\cite{nishi,hn98,wada} 
which may be characterized by the soliton equation.\cite{komatsu}
We call a front interface the kink  and a backward interface the
antikink. The antikink is not stable in the actual
situations and emits dispersive waves backward. The waves are caught
by the next domain.
In the simplest situation, we can ignore the
widths of the kinks and antikinks which may be much smaller than the typical
domain size.

From the observation of experiments
 the power spectrum  may not be related to the formation process of domains
but be characterized by the emission of
dispersive waves from an antikink. Thus, we ignore the formation of 
a congested domain but focus on the decay process of the domain. 
We also map the
model onto a one-dimensional space, where the position fixed in an
experimental system is denoted by $x$ and the
system size is  $L$ and the boundaries are located at $x=\pm L/2$.  
For simplicity, we
place a detector to measure the power spectrum at $x=0$, {\it i.e.} the
center of the system.
Let us introduce
the packing fraction $\phi(x,t)\equiv n(x,t)/n_0$ where $n_0$ is the
maximum density.

If we assume that an idealistic congested domain exists in the system at 
time $t=0$, 
the packing fraction is given by 
$\phi(x,t=0)=1$ between $x=x_0$ and $x=x_0+l$, 
and $\phi(x,0)=0$ for otherwise, where $l$ and $x_0$ are the size of the
domain and the position of an antikink at $t=0$, respectively. 
The equivalent expression  is 
\begin{eqnarray}\label{1}
\phi(x,0)&=&\frac{l}{L}+\sum_{n=1}^{\infty}\frac{\cos \frac{2n\pi x}{L}}{n\pi}\{\sin\frac{2n \pi(x_0+l)}{L}-\sin\frac{2n\pi x_0}{L}\} \nonumber \\
&&-\sum_{n=1}^{\infty}\frac{\sin \frac{2n\pi x}{L}}{n\pi}\{\cos\frac{2n \pi(x_0+l)}{L}-\cos\frac{2n\pi x_0}{L}\} .
\end{eqnarray}
On the other hand, the antikink is
unstable because of the dispersion of propagating velocity, though we
can ignore such the effect for the kink. Thus, we assume that the time
dependence of $\phi(0,t)$ can be described by
\begin{equation}\label{3}  
\phi(0,t)=\frac{l}{L}+\sum_{n=1}^{\infty}\frac{1}{n\pi}
\{\sin \frac{2n\pi(x_0+l+c_0 t)}{L}
-\sin\frac{2n\pi(x_0+c_0t(1-(\frac{2n\pi \xi}{L})^2))}{L}
\} ,
\end{equation}
where $c_0$ is the average speed of domains
and $\xi$ is a characteristic length scale
of the dispersion relation. We note that the shape of domain at time $t$ 
is no longer idealistic one but is decayed. 

With the aid of Wiener-Khinchin theorem, the power spectrum $I(\omega)$ 
and the auto-correlation function $C(t)$  can be written as
\begin{equation}\label{2}
I(\omega)\equiv \frac{1}{\sqrt{2\pi}}\int_{-\infty}^{\infty} dt e^{i \omega t} C(t), \qquad C(t)\equiv <\phi(0,0)\phi(0,t)> ,
\end{equation}
where the ensemble average in eq.(\ref{2}) is interpreted as the
average by the position of the antikink $x_0$. Because the domain
propagates with the constant speed $c_0$ if we neglect the dispersion,
the existence probability of  domains should be uniform except for the
boundary regions. Thus, we may assume the probability
distribution function $P(x_0)=1/L$ and  $C(t)=
\frac{1}{L}\int_{-L/2}^{L/2}dx_0\phi(0,0)\phi(0,t) $.
Substituting eqs.({\ref{1}}) and (\ref{3}) into (\ref{2}) we obtain
\begin{equation}
C(t)= \frac{l^2}{L^2}+J_0(t)
+J_1(t)+J_2(t) ,
\label{4}\end{equation}
where
\begin{eqnarray}
J_0(t)&=&\sum_{n=1}^{\infty}\frac{1}{2\pi^2n^2}\{\cos\frac{2\pi n c_0 t}{L}
-\cos\frac{2n\pi}{L}(l+c_0t)\} \\
J_1(t)&=-&\sum_{n=1}^{\infty}\frac{1}{2n^2\pi^2}\{
1-\cos\left[\frac{2\pi n}{L}c_0t\right]
\cos\left[(\frac{2\pi n}{L})^3\xi^2c_0t\right]
-\sin\left[\frac{2\pi n}{L}c_0t\right]
\sin\left[(\frac{2\pi n}{L})^3\xi^2c_0t\right]\}
\label{5} \\
J_2(t)&=&\sum_{n=1}^{\infty}\frac{1}{2n^2\pi^2}\{
1-\cos\left[\frac{2\pi n}{L}(l-c_0t)\right]\cos\left[(\frac{2\pi n}{L})^3\xi^2c_0t\right] \nonumber\\
& &+\sin\left[\frac{2\pi n}{L}(l-c_0t)\right]\sin\left[(\frac{2\pi n}{L})^3\xi^2c_0t\right]\}
 .\label{6}
\end{eqnarray}
Here, 
$J_0(t)$  in eq.(\ref{4}) can be calculated as
\begin{equation}
J_0(t)
=\frac{l}{2L}(1-\frac{l}{L})-\frac{l c_0t}{L^2},
\label{7}\end{equation}
where we use the formula $\sum_{n=1}^{\infty}\cos nx/n^2=\pi^2/6-\pi
x/2+x^2/4$. Thus,
$I_0(\omega)\equiv\frac{1}{\sqrt{2\pi}}\int_{-\infty}^{\infty}dt e^{i
\omega t}(l^2/L^2+J_0(t))$ becomes
\begin{equation}\label{I0}
I_0(\omega)=\sqrt{2\pi}\frac{l(L+l)}{2L^2}\delta(\omega)+\displaystyle\sqrt{\frac{2}{\pi}}\frac{l c_0}{L^2}\omega^{-2}.
\end{equation}

The evaluations of $J_1(t)$ and $J_2(t)$ are nontrivial.
When we assume $c_0t\ll L$ the summation in
$J_1(t)$  can be replaced by the integral. From the expansion
  by $c_0t/\xi$
we obtain
\begin{eqnarray}
J_1(t)&\simeq& -\frac{1}{3\pi L}(\xi^2c_0t)^{1/3}
[
\int_0^{\infty}dz\frac{1-\cos z}{z^{4/3}}-
\left(\frac{c_0t}{\xi}\right)^{2/3}\int_0^{\infty}dz \frac{\sin z}{z}
] 
\nonumber\\
&=& -\frac{1}{3 L\Gamma(4/3)}(\xi^2c_0t)^{1/3}+\frac{c_0t}{6L},
\label{8}
\end{eqnarray}
where we use $\int_0^{\infty}dz(1-\cos z)/z^{4/3}=\pi/\Gamma(4/3)$ and
$\int_0^{\infty}dz \sin z/z=\pi/2$ with the Gamma function $\Gamma(z)$.
The corresponding Fourier transform of $J_1(t)$ is thus given by
\begin{equation}\label{I1}
I_1(\omega)=\frac{\sqrt{2}}{6\sqrt{\pi}L}(\xi^2c_0)^{1/3}\omega^{-4/3}
-\frac{\sqrt{2}c_0}{6\sqrt{\pi}L}\omega^{-2}.
\end{equation}

On the other hand, for $l\gg c_0 t$, $J_2(t)$ can be evaluated as
\begin{eqnarray}
J_2(t)&\simeq &\sum_{n=1}^{\infty}\frac{1}{2\pi^2n^2}\{
1-\cos\frac{2\pi n l}{L}\cos(\frac{2\pi n}{L})^3\xi^2c_0t
+\sin\frac{2\pi n l}{L}\sin(\frac{2\pi n}{L})^3\xi^2c_0t \nonumber\\
& &-\frac{2\pi n}{L}c_0t \cos\frac{2\pi n l}{L}
\sin[\left(\frac{2\pi n}{L}\right)^3\xi^2c_0t]\} \nonumber\\
&\simeq& \frac{l}{\pi L}J_{21}(t)-\frac{c_0 t}{\pi L}J_{22}(t).
\label{9}
\end{eqnarray}
Here $J_{21}(t)$ and $J_{22}(t)$ are respectively given by\cite{math}
\begin{eqnarray}
J_{21}(t)&\equiv&\int_0^{\infty}dx\frac{1}{x^2}
(1-\cos x\cos[x^3b t]+\sin x\sin[x^3 bt]) \nonumber \\
&=& \frac{1}{720}[-120\sqrt{3}(b t)^{1/3}\Gamma(-1/3)
{}_{1}F_4[-\frac{1}{6};\frac{1}{6},\frac{1}{2},\frac{2}{3},\frac{5}{6};\frac{1}{11664 (b t)^2}] \nonumber \\
& &+ 60\sqrt{3}(b t)^{-1/3}\Gamma(1/3)
{}_1F_4[\frac{1}{6};\frac{1}{2},\frac{5}{6},\frac{7}{6},\frac{4}{3};\frac{1}{11664 (b t)^2}] \nonumber \\
& &+\{120\pi-20\sqrt{3}(b t)^{-2/3}\Gamma(2/3)
{}_1F_4[\frac{1}{3};\frac{2}{3},\frac{7}{6},\frac{4}{3},\frac{3}{2};\frac{1}{11664 (b t)^2}] \nonumber \\
& &+\sqrt{3}(b t)^{-4/3}\Gamma(4/3)
{}_1F_4[\frac{2}{3};\frac{4}{3},\frac{3}{2},\frac{5}{3},\frac{11}{6};\frac{1}{11664 (b t)^2}]
\}
]\label{10}
\end{eqnarray}
with $b\equiv \xi^2 c_0/l^3$, and
\begin{eqnarray}
J_{22}(t)&\equiv&
\int_0^{\infty}
dx\frac{\cos x\sin[x^3b t]}{x} \nonumber \\
&=&\frac{1}{216(b t)^{4/3}}
[- 18\sqrt{3}\Gamma(2/3)(b t)^{2/3}{}_1F_4[\frac{1}{3};\frac{1}{2},\frac{2}{3},\frac{7}{6},\frac{4}{3};\frac{1}{11664 (b t)^2}] \nonumber\\
& &+\pi\{36(b t)^{4/3}+\frac{1}{\Gamma(4/3)}{}_1F_4[\frac{2}{3};\frac{5}{6},\frac{4}{3},\frac{3}{2},\frac{5}{3};\frac{1}{11664 (b t)^2}\}
], \label{11}
\end{eqnarray}
where
${}_1F_4[a_1;b_1,b_2,b_3,b_4;z]\equiv\sum_{k=0}^{\infty}(a_1)_kz^k/[(b_1)_k\cdots
(b_4)_kk!]$ with $(\alpha)_k=\alpha(\alpha+1)\cdots(\alpha+k-1)$ 
is the generalized hypergeometric function.

The expressions of $J_{21}(t)$ and $J_{22}(t)$ are complicated. For example,
the Fourier transforms of $J_{21}(t)$ is given by\cite{math}
\begin{eqnarray}
I_{21}(\omega)&=&-\frac{\sqrt{\pi/2}}{360b^{4/3}\omega^{4/3}}
(120 b^{5/3}{}_0F_5[;\frac{1}{6},\frac{1}{3},\frac{1}{2},\frac{2}{3},\frac{5}{6};-\frac{\omega^2}{46656b^2}]\nonumber\\
& &-60b\omega^{2/3}{}_0F_5[;\frac{1}{2},\frac{2}{3},\frac{5}{6},\frac{7}{6},\frac{4}{3};-\frac{\omega^2}{46656b^2}]\nonumber\\
& &+20\sqrt{3}b^{2/3}\omega {}_0F_5[;\frac{2}{3},\frac{5}{6},\frac{7}{6},\frac{4}{3},\frac{3}{2};-\frac{\omega^2}{46656b^2}]\nonumber\\
& &-10b^{1/3}\omega^{4/3}{}_0F_5[;\frac{5}{6},\frac{7}{6},\frac{4}{3},\frac{3}{2},\frac{5}{3};-\frac{\omega^2}{46656b^2}]\nonumber\\
& &+\sqrt{3}\omega^{5/3}{}_0F_5[;\frac{7}{6},\frac{4}{3},\frac{3}{2},\frac{5}{3},\frac{11}{6};-\frac{\omega^2}{46656b^2}]
) . 
\label{power-I21}\end{eqnarray}
It should be noted that the Fourier transform of $J_{22}(t)$ becomes 
complex. This is because the expression we obtain depends on the
initial condition and the choice of the frame.
%the imaginary part represents the change of the frame. 
If the system is Galilei invariant, such the term should be
zero. Therefore, we regard the contribution from $J_{22}(t)$ as
unimportant. 

In the limit of
$\omega\to 0$, $I_2(\omega)$ becomes 
\begin{equation}
I_2(\omega)\simeq \frac{l}{\pi L}I_{21}(\omega)\to 
-\frac{\sqrt{2}}{6\sqrt{\pi}L}(\xi^2c_0)^{1/3}\omega^{-4/3}.
\label{12}\end{equation}
%and
%\begin{equation}
%I_{22}(\omega)\to i \frac{\sqrt{2\pi}}{9b^{1/3}}\omega^{-5/3} 
%\label{13}\end{equation}
It is notable that this asymptotic expression
of $I_2(\omega)$ is canceled with  the term proportional to
$\omega^{-4/3}$ in
$I_1(\omega)$. That is, the spectrum obeying $\omega^{-4/3}$ disappears
and $I(\omega)\sim \omega^{-2}$ in the limit of $\omega\to 0$. 
%for $\omega\to 0$, 
%where $I_{21}(\omega)$ and $I_{22}(\omega)$
%are respectively the Fourier Transform of $J_{21}(t)$ and $t J_{22}(t)$. 

On the other hand, though $I_{21}(\omega)$ is singular in the limit of
 $\omega\to\infty$, 
$I_{21}(\omega)$ is regular for enough large $\omega$. In fact,
one can obtain the analytic
expansion of $J_{21}(t)$  near $bt=0.001$ as $2J_{21}(t)/\pi\simeq
1+\alpha(bt-0.001)+O((bt-0.001)^2)\simeq 1+\alpha b t+\cdots$ 
with $\alpha=0.000229538$.
If we replace $J_{21}(t)$ by this approximate function, we obtain the
approximate Fourier transform
\begin{equation}
I_{21}(\omega)\sim \frac{\pi^{3/2}}{\sqrt{2}}\delta(\omega)-\displaystyle\sqrt{\frac{\pi}{2}}\alpha b\omega^{-2}
\end{equation}
for large $\omega$.

Thus, we obtain the power spectrum 
$I(\omega)=I_0(\omega)+I_1(\omega)+I_2(\omega)$ as
\begin{equation}\label{eq1}
I(\omega)=\frac{\sqrt{2}}{6\sqrt{\pi}L}(\xi^2c_0)^{1/3}\omega^{-4/3}
-\frac{\sqrt{2}c_0}{6\sqrt{\pi}L}(1-\frac{6l}{L})\omega^{-2}+\frac{l}{\pi L}I_{21}(\omega)
\end{equation}
for $\omega\ne 0$. 
For large $\omega$,  $I(\omega)$ is dominated by
 the term proportional to
$\omega^{-4/3}$ as
\begin{equation}\label{eq2}
I(\omega)\to \frac{\sqrt{2}}{6\sqrt{\pi}L}(\xi^2c_0)^{1/3}\omega^{-4/3}.
\end{equation}
Thus, we  derive the spectrum obeying
$\omega^{-4/3}$. Figure 2 shows the comparison of eq.(\ref{eq1}) with
eq.(\ref{eq2}), where we can see the tail obeying $\omega^{-4/3}$ for large 
$\omega$, while eq.(\ref{eq1}) seems to obey $\omega^{-2}$ for small $\omega$. 
It is obvious that both expressions (\ref{eq1}) and (\ref{eq2}) become
identical for larger $\omega$.

It should be noted that the actual process includes many other factors
for larger $\omega$ and smaller $\omega$. In experiments, $I(\omega)$
decays exponentially for larger $\omega$, because the initial state is
not in an idealistic  domain as  we have assumed here. To reproduce the
full shape of spectrum we need to contain the formation process of
domains for our analysis. That is an important future problem to be
solved. Nevertheless, we
believe that our picture presented here captures the essence of physics and clarify the 
mechanism of emergence of $\omega^{-4/3}$ law.

In this Communication, we have demonstrated that
 the main process to produce $\omega^{-4/3}$ law is the
emission of the dispersive wave from an antikink. 
This result is universal when isolated congested domains exist in
 a dissipative flow.
Through the analysis,
 we have revised the previous uncertain picture. 

The author thanks Namiko Mitarai for fruitful discussion.
This study is partially supported by the Grants-in-Aid of Japan Space
Forum, and
Ministry of Education, Culture, Sports, Science and Technology (MEXT), 
Japan (Grant No. 15540393) and the Grant-in-Aid for the 21st century COE 
"Center for Diversity and Universality in Physics" from MEXT, Japan.

%%%%%%%%%%%%%%%%%%%%%%%%%%%%%%
\begin{figure}
%[htbp]
\epsfxsize=10cm
\centerline{\epsfbox{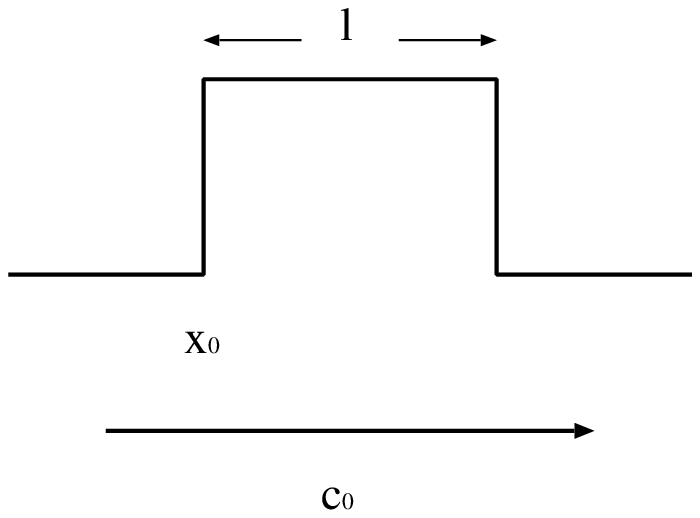}}
% \begin{center}
%  \includegraphics[width=100mm]{fig1.eps}
% \end{center}
\caption{
A schematic picture of the propagation of a domain (its size $l$) with 
the speed $c_0$. The dispersive wave emits from the antikink interface at 
 $x_0$.
} 
\end{figure}
%%%%%%%%%%%%%%%%%%%%%%%
%%%%%%%%%%%%%%%%%%%%%%%%%%%%%%
\begin{figure}
%[htbp]
\epsfxsize=14cm
\centerline{\epsfbox{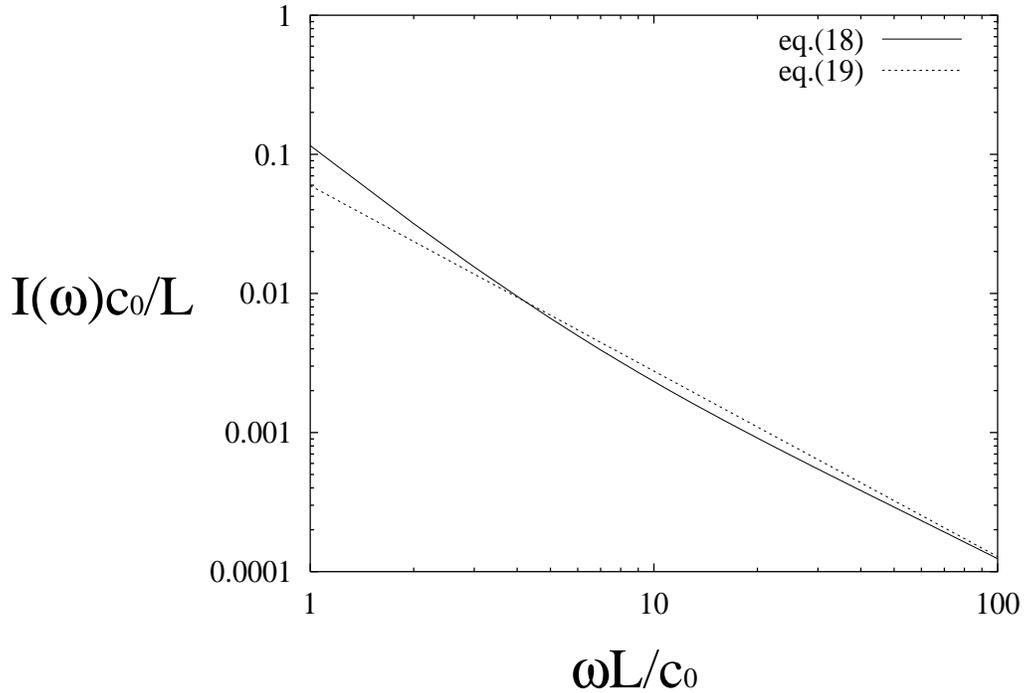}}
% \begin{center}
%  \includegraphics[width=140mm]{Fig2.eps}
% \end{center}
\caption{
Log-log plots of eqs.(\ref{eq1}) and (\ref{eq2}) as the frequency spectra. We
 adopt the parameters $l/L=\xi/L=0.3$ and $L b/c_0=L/l=10/3$.
.} 
\end{figure}
%%%%%%%%%%%%%%%%%%%%%%%

\begin{references}
\bibitem{math} We have used MATHEMATICA for the calculation.
\end{references}
\end{document}